\documentclass[a4paper,12pt]{article}
\usepackage[dvips]{graphicx}
\usepackage{amsmath}
\begin{document}
\begin{center}
{\large \bf On a problem of dark matter: A centrally symmetric distribution of a pseudoscalar field in the general relativity}
\vspace{5ex}

I. K. Rozgacheva$^a$, A. A. Agapov$^b$
\vspace{2ex}

{\small \it Russian Institute for Scientific and Technical Information (VINITI RAS), Moscow, Russia}
\vspace{1ex}

E-mail: $^a$rozgacheva@yandex.ru, $^b$agapov.87@mail.ru
\vspace{4ex}
\end{center}
\begin{abstract}
A precise solution of the general relativity equations for centrally symmetric distribution of a pseudoscalar field with U(1) symmetry is presented. It is found that energy density of the field is restricted in the symmetry center and falls with radial distance at a much slower rate than in numerical simulations of dark matter haloes. This result can solve the cusp problem for galaxies and the problem of rotation curves for galaxy exteriors.
\end{abstract}
\vspace{1ex}

\vspace{3ex}

\begin{center}
{\bf 1. Introduction}
\end{center}

The dark matter model consisting of hypothetic pseudoscalar particles - axions - is used in the $\Lambda$CDM cosmological model. Their speed is assumed to be small, therefore such dark matter may be considered as cold dark matter \cite{1}. Simulations of the large-scale structure formation in the framework of $\Lambda$CDM can achieve a satisfying agreement with observational data for large spatial scales. However there are some contradictions on small scales such as galaxies and their groups. One of them is related to the gravitating matter density distribution. According to the observational data, in central regions of galaxies with low surface brightness this density almost does not depend on radial distance and keep to be a restricted value. On the other hand, in simulations of such galaxies formation in dark matter haloes their density is found to fall inversely proportionally to radial distance. Such dependence of density is called "cusp". One of possible solutions of this cusp-absence problem is proposed in \cite{2}, where the role of small-scale density fluctuations and their velocities is taken into account (there is a constraint on small scales in computer simulations). These small-scale motions of dark matter can "wash out" cusps.

This work presents a precise solution of general relativity equations for centrally symmetric distribution of a pseudoscalar field with U(1) symmetry. It is found that energy density of the field is restricted in the symmetry center and falls along radial distance at a much slower rate than in numerical simulations of dark matter haloes. This result shows that the pseudoscalar field allows solving not only the cusp problem but also the problem of rotation curves for galaxy exteriors.
\vspace{2ex}

\begin{center}
{\bf 2. Solution of Einstein's and Lagrange's equations for a pseudoscalar field with U(1) symmetry}
\end{center}

In our works \cite{3}, \cite{4} we find a solution of the Hilbert-Einstein equation and the Lagrange equation for a pseudoscalar field (a charged scalar meson field of matter) $\psi$ with rotary symmetry
$$
\psi\psi^*=\Psi^2=\mbox{const}, \eqno (1)
$$
where the asterisk denotes complex conjugation and $\Psi$ is the field amplitude relating to the field charge $Q\sim\Psi^2$.

Let's consider the Einstein-Hilbert action:
$$
S=-\frac{c^3}{16\pi G}\int\left(R-\frac{8\pi G}{c^4}L\right)\sqrt{-g}\,d^4x,
$$
where $R$ is the scalar curvature, $g<0$ is a determinant of the metric tensor $g_{mn}$, the space-time interval is $ds^2=g_{mn}dx^mdx^n$, the indices take values 0, 1, 2, 3, the metric signature is $(+---)$. We use the following form of the complex field Lagrangian:
$$
L=\frac1{hc}\left(g^{mn}\frac{\partial\psi}{\partial x^m}\frac{\partial\psi^*}{\partial x^n}-U\left(\psi\psi^*\right)\right),\eqno (2)
$$
where $U\left(\psi\right)$ is the field potential, $h$ is Planck's constant, $c$ is the light velocity. Hereafter, the field dimension is $\left[\psi\right]=\mbox{erg}$, the contravariant metric tensor dimension is $\left[g^{mn}\right]=\mbox{cm}^{-2}$. This field possesses the symmetry (1). Its Lagrange equation is
$$
\frac1{\sqrt{-g}}\frac{\partial}{\partial x^n}\left(\sqrt{-g}g^{mn}\frac{\partial\psi}{\partial x^m}\right)=-\frac{\partial U}{\partial\psi^*}.\eqno (3)
$$

In Einstein's equation
$$
R_n^m-\frac12R\delta_n^m=\kappa T_n^m\eqno(4)
$$
the energy-momentum tensor of the complex field equals
$$
T_n^m=\frac{\partial\psi}{\partial x^n}\frac{\partial L}{\partial \left(\frac{\partial\psi}{\partial x^m}\right)}+\frac{\partial\psi^*}{\partial x^n}\frac{\partial L}{\partial \left(\frac{\partial\psi^*}{\partial x^m}\right)}-\delta_n^mL=\frac1{hc}g^{mp}\left(\frac{\partial\psi}{\partial x^p}\frac{\partial\psi^*}{\partial x^n}+\frac{\partial\psi}{\partial x^n}\frac{\partial\psi^*}{\partial x^p}\right)-\delta_n^mL,
$$
where $R_n^m$ is the Ricci tensor, $\kappa=8\pi G/c^4$ is Einstein's gravity constant, $G$ is Newton's gravity constant, $\delta_n^m$ is the delta symbol.

Following form of potential is used further:
$$
U=U_0\psi\psi^*.\eqno (5)
$$
One can ascertain that Lagrange's equation (3) with potential (5) is satisfied for the solution:
$$
\psi =\Psi\mbox{e}^{i\varphi},\ \ \ \psi^*=\Psi\mbox{e}^{-i\varphi},
$$
$$
\Gamma_{mn}^l=\frac1{U_0}\frac{\partial^2\varphi}{\partial x^m\partial x^n}\left(g^{lp}\frac{\partial\varphi}{\partial x^p}+a^l\right),\eqno (6)
$$
$$
g_{mn}=\frac1{U_0}\left(4\frac{\partial\varphi}{\partial x^m}\frac{\partial\varphi}{\partial x^n}+\frac{\partial\varphi}{\partial x^m}a_n+\frac{\partial\varphi}{\partial x^n}a_m\right),
$$
$$
T_{mn}=\frac{2\Psi^2}{hc}\frac{\partial\varphi}{\partial x^m}\frac{\partial\varphi}{\partial x^n}.
$$
where the field phase $\varphi\left(x^m\right)$ is a differentiable function. Hereafter, indices are raised and lowered with the metric tensor, indices appearing twice in a single term imply summing over its values, semicolon denotes covariant differentiation, $\displaystyle \Gamma_{mn}^l$ are the Christoffel symbols.

Derivative $\displaystyle \frac{\partial\varphi}{\partial x^m}$ and covariant vector $a_m$ satisfy the equations:
$$
g^{mn}\frac{\partial\varphi}{\partial x^m}\frac{\partial\varphi}{\partial x^n}=U_0,
$$
$$
g^{mn}\left(\frac{\partial\varphi}{\partial x^m}\right)_{;n}=0,
$$
$$
a_{m;l}=0, \eqno (7)
$$
$$
a_ma^m=-3U_0,
$$
$$
\frac{\partial\varphi}{\partial x^m}a^m=0.
$$
Covariant $a_m$ and contravariant $a^k$ vectors satisfy the equations:
$$
\frac{\partial a_m}{\partial x^l}=-3\frac{\partial^2\varphi}{\partial x^m\partial x^l},
$$
$$
\frac{\partial a^n}{\partial x^m}a_n=3a^n\frac{\partial^2\varphi}{\partial x^n\partial x^m}.\eqno (8)
$$
One can ascertain through a substitution that the following equalities are satisfied for the solution (6-7):
$$
\frac{\partial g_{mn}}{\partial x^l}=g_{km}\Gamma_{nl}^k+g_{kn}\Gamma_{ml}^k,
$$
$$
\Gamma_{kl}^m=\frac12g^{mn}\left(\frac{\partial g_{nk}}{\partial x^l}+\frac{\partial g_{nl}}{\partial x^k}-\frac{\partial g_{kl}}{\partial x^n}\right),
$$
$$
\delta_m^n=g^{nl}g_{lm}.
$$

One can ascertain through a substitution that solutions (6) satisfy the Einstein-Hilbert equation (4).

The phase path of the fields $\psi$ and $\psi^*$ is the circle (1):
$$
\psi\psi^*={\psi_1}^2+{\psi_2}^2=\Psi^2,
$$
$$
\psi=\psi_1+i\psi_2,\ \ \ \psi^*=\psi_1-i\psi_2.
$$
The function $\varphi$ is degree of rotation round the circle. Length of a circle arc i.e. interval of set $\left\{\psi_1,\psi_2\right\}$ equals
$$
dF^2=\left(d\psi_1\right)^2+\left(d\psi_2\right)^2=d\psi d\psi^*=\Psi^2\frac{\partial\varphi}{\partial x^m}\frac{\partial\varphi}{\partial x^n}dx^mdx^n. \eqno (9)
$$

Note that a motion along the phase corresponds to a massless field - a Goldstone boson. The expression (9) shows that the solution (6) describes the geometry of this boson.

The cosmological model corresponding to the solution (6-8) is described in our works \cite{3}, \cite{4} and its cosmological constraints are proposed in our work \cite{5}.

\begin{center}
{\bf 3. Solution for a centrally symmetric case}
\end{center}

Let the phase $\varphi$ is a function of an argument $y=x^nk_n$, where $k_{n;m}=0$ and $a^ik_i=0$. Let us consider a pseudoscalar field whose gravitational field possesses a spatial central symmetry. In this case the space-time coordinates may be chosen in such a way that the space-time interval has a form
$$
ds^2=g_{00}dx^0dx^0+g_{11}dx^1dx^1+g_{22}dx^2dx^2+g_{33}dx^3dx^3.
$$

The non-zero components of the metric tensor are equal to
$$
g_{00}=\frac2{U_0}\frac{d\varphi}{dy}\left(2\frac{d\varphi}{dy}+\frac{a_0}{k_0}\right)k_0k_0,
$$
$$
g_{11}=\frac2{U_0}\frac{d\varphi}{dy}\left(2\frac{d\varphi}{dy}+\frac{a_1}{k_1}\right)k_1k_1,
$$
$$
g_{22}=\frac2{U_0}\frac{d\varphi}{dy}\left(2\frac{d\varphi}{dy}+\frac{a_2}{k_2}\right)k_2k_2,
$$
$$
g_{33}=\frac2{U_0}\frac{d\varphi}{dy}\left(2\frac{d\varphi}{dy}+\frac{a_3}{k_3}\right)k_3k_3.
$$

Let us consider a solution satisfying the equation
$$
\frac{\partial^2\varphi}{\partial x^i\partial x^m}=b_i\frac{\partial\varphi}{\partial x^m}+b_m\frac{\partial\varphi}{\partial x^i},
$$
where the constant components of the vector $b_i$ satisfy the condition $b_ik^i=0$. In this case the equations (4) comes to the following equations:
$$
\left(\frac{d\varphi}{dy}\right)^2k^2=U_0, \eqno(10)
$$
$$
k^2\frac{d\varphi}{dy}\frac{da^i}{dy}b_i=\frac{2\kappa}{hc}\Psi^2U_0^2-\left(a^ib_i\right)^2.
$$

Without loss of generality let's accept that $k=const$ and $b_i$ are constant components of a 4-vector. Then the equations (10) have the following solution:
$$
\frac{d\varphi}{dy}=\frac{\sqrt{U_0}}{k}, \eqno(11)
$$
$$
a^ib_i=\left(a^ib_i\right)_*+\left(\frac{2\kappa}{hc}\frac{\Psi^2U_0}{k^2}\right)^{1/2}\tanh\left[\left(\frac{2\kappa}{hc}\frac{\Psi^2U_0}{k^2}\right)^{1/2}y\right],
$$
where $\left(a^ib_i\right)_*$ is an integration constant.

The solution (11) is restricted:
$$
a^ib_i\le\left(a^ib_i\right)_*+\left(\frac{2\kappa}{hc}\frac{\Psi^2U_0}{k^2}\right)^{1/2}
$$
and its 8 free parameters $b_i$ and $k_i$ are related by two conditions: $b_ik^i=0$ and $k_ik^i=const$.

The energy density of the pseudoscalar field equals
$$
\displaystyle T_0^0=g^{00}T_{00}=\frac{\Psi^2U_0}{2hc}\frac1{\displaystyle 1+\frac1{2\sqrt{U_0}}\frac{ka_0}{k_0}}. \eqno(12)
$$

It follows from the conditions $a_ma^m=-3U_0$ and $k=const$ that there is a class of solutions with $a^m\sim1/a_m$ and $k_m\sim1/k^m$. For such a class of solutions the energy density of the pseudoscalar field (12) is a restricted value. For example, if the parameters $b_i$ and $k_i$ are chosen in such a way that
$$
\frac{a_0}{k_0}=\frac{k^0}{a^0}=\frac{k^0b_0}{a^0b_0}=\frac{k^0b_0}{\displaystyle \left(a^ib_i\right)_*+\left(\frac{2\kappa}{hc}\frac{\Psi^2U_0}{k^2}\right)^{1/2}\tanh\left[\left(\frac{2\kappa}{hc}\frac{\Psi^2U_0}{k^2}\right)^{1/2}y\right]},
$$
than the energy density of the pseudoscalar field equals
$$
T_0^0=\frac{\Psi^2U_0}{4hc}
\left[
\left(1+\frac{\left(a^ib_i\right)_*}{\displaystyle \left(\frac{2\kappa}{hc}\frac{\Psi^2U_0}{k^2}\right)^{1/2}}\right)-
\left(1-\frac{\left(a^ib_i\right)_*}{\displaystyle \left(\frac{2\kappa}{hc}\frac{\Psi^2U_0}{k^2}\right)^{1/2}}\right)
e^{-2\left(\frac{2\kappa}{hc}\frac{\Psi^2U_0}{k^2}\right)^{1/2}y}
\right], \eqno(13)
$$
for
$$
k^0b_0=\frac{\Psi^2U_0}{k^2}\left(\frac{2\kappa}{hc}\right)^{1/2}+\frac{2\sqrt{U_0}}{k}\left(a^ib_i\right)_*.
$$

The solution (11-13) describes the centrally symmetric distribution of a pseudoscalar field energy density which is restricted in the center (i.e. there is no cusp) and falls with radial distance tending to a restricted constant value. Such energy density distribution is analogous to the Einasto profile \cite{6} for the density distribution in galaxies.

\end{document}